\documentclass[12pt,twoside]{article}
\begin{document}
\baselineskip=0.20in
\vspace{20mm}
\baselineskip=0.30in
\begin{center}

{\large \bf Relativistic Treatment of the Spin-Zero Particles Subject to the 
q-Deformed Hyperbolic Modified P\"{o}schl-Teller Potential.}
\vspace{10mm}

{ \large  K.\,J. Oyewumi\footnote{E-Mail:~~kjoyewumi66@unilorin.edu.ng} and T.\,T. Ibrahim\footnote{Department of Physics, University of Stellenbosch, Matieland, PO 1529, Stellenbosch, 7599, South     
  Africa.   
} } \vspace{5mm}

 { Theoretical Physics Section, Department of Physics\\ University of Ilorin,  P. M. B. 1515, Ilorin, Nigeria. \\}

{\large   S.\, O. Ajibola} 

{
Department of Mathematics, Faculty of Science and Technology,  \\National Open University of Nigeria, Lagos, Nigeria.
}

{\large  D.\, A. Ajadi }

{
Department of Physics, Ladoke Akintola University of Technology, \\Ogbomoso, Nigeria. 
}

\baselineskip=0.20in

\vspace{4mm}

\end{center}


\begin{abstract}
\noindent
In this study, we solve the Klein-Gordon equation with equal scalar and vector q-deformed hyperbolic modified P\"{o}schl-Teller potential. The explicit expressions of bound state spectra and the normalized eigenfunctions for s-wave bound states are obtained analytically. The energy equations and the corresponding wave functions for the special cases of the equally mixed q-deformed hyperbolic  modified P\"{o}schl-Teller potential for spinless particle are briefly discussed.

\end{abstract}

\baselineskip=0.28in \vspace{2mm}


\baselineskip=0.28in \vspace{2mm}

 \vspace{2mm} 

\baselineskip=0.28in \vspace{2mm}

\noindent
\section{\bf Introduction}
In recent years, there has been increasing interest in finding the analytical solutions of relativistic and non-relativistic quantum mechanical problems. The presence of strong fields or high speeds introduces relativistic phenomena that cannot be described using the Schr\"{o}dinger equation. In relativistic quantum mechanics, the solutions of the Klein-Gordon and Dirac equations with various physical potential play an important role in nuclear physics and other related areas. The Klein-Gordon equation has also been used to understand the motion of the spin-zero particles in various potentials. 

Many studies have been carried out to explore the relativistic energy eigenvalues and the corresponding wave functions by solving the Klein-Gordon and Dirac equations (Greiner 2000, Simsek and E$\check{g}$rifes 2004, Qiang 2004, Chen 2005, Zhao et al. 2005, Alhaidari et al. 2006, Berkdermir 2007, Durmus and Yasuk 2007, Qiang et al. 2007, Soylu et al. 2008, Liu et al. 2009, Motavalli 2009 and Xu et al. 2010, Oyewumi 2010).  For analytic solutions of these equations of motion with various potentials, some authors have assumed equal scalar and vector potentials (see Alhaidari et al. 2006 for details). 

Several methods ranging from exact analytical technique to approximate analytical methods have also been used in solving the relativistic quantum mechanical problems with different potentials.
The generalized symmetrical double-well potential (Zhao et al. 2005); Kratzer-type and generalized ring-shaped Kratzer potentials (Qiang 2004, Berkdermir 2007, Oyewumi 2010); exponential-type potentials (Simsek and E$\check{g}$rifes 2004); double ring-shaped Kratzer potential (Durmus and Yasuk 2007); Hulth$\acute{e}$n (Qiang et al. 2004, Guo et al. 2003); ring-shaped harmonic oscillator (Qiang 2003); Rosen-Morse-type potential (Yi et al. 2004, Soylu et al. 2008); Eckart potential (Zou et al. 2005, Dong et al. 2007, Liu et al. 2009, Yahya et al. 2010); Manning-Rosen potential (Qiang and Dong 2007) and Scarf-type potential (Zhang et al. 2005, Motavalli 2009)  are examples of potentials which finds application in various aspects of modern physics. These have been studied in both the relativistic and non-relativistic limits.

In this paper, we study analytical bound state solutions of the Klein-Gordon equation with the equal scalar and vector q-deformed hyperbolic modified P\"{o}schl-Teller potential.  P\"{o}schl-Teller potential (P\"{o}schl and Teller 1933, Landau and Liftshitz 1977) is found very useful in the study of the $\Lambda$-hypernuclear in nuclear physics and other related areas (Oyewumi and Bangudu 1999, 2000, Grypeos et al. 2004, Oyewumi et al. 2004, Efthymiou et al. 2008, Oyewumi 2009). 

We have also considered the special cases of this potential in the Klein-Gordon equation, the bound state energy spectra and the corresponding wave functions of the resulting potentials which include the reflectionless-type potential, q-deformed symmetric hyperbolic modified P\"{o}schl-Teller potential, symmetric modified P\"{o}schl-Teller potential and the PT-symmetric version of the  hyperbolic modified P\"{o}schl-Teller potential are obtained.

	This paper is organized as follows. Section $2$ contains the bound state solutions of the q-deformed hyperbolic modified P\"{o}schl-Teller potential. In Section $3$, we investigate the special cases of this potential and finally, conclusion is given in Section $4$.

\section{Bound State Solutions of the q-Deformed Hyperbolic Modified P\"{o}schl-Teller Potential.}

The time-independent three-dimensional Klein-Gordon equation with the scalar potential $V_{s}(r)$ and vector potential $V_{v}(r)$ is given by (Greiner 2000, Simsek and E$\check{g}$rifes 2004, Qiang 2004, Chen 2005, Zhao et al. 2005, Alhaidari et al. 2006, Berkdermir 2007, Durmus and Yasuk 2007, Qiang et al. 2007, Soylu et al. 2008, Liu et al. 2009, Motavalli 2009 and Xu et al. 2010)
\begin{equation}
\left[\hbar^{2} c^{2} \nabla^{2}  + \left(\mu c^{2} + \frac{V_{s}(r)}{2}\right)^{2} - \left(E_{R} - \frac{V_{r}(r)}{2}\right)^{2} \right]\Psi(r, \theta, \phi) = 0
\label{r1},
\end{equation}
where $E_{R}$ is the relativistic energy of the system and $\mu$ denotes the rest mass of particle.

	For spherically symmetric scalar and vector potentials and by putting 
\begin{equation}
\Psi_{n,~\ell,~m}(r, \theta, \phi) = \frac{U_{n,~\ell}(r) }{2} Y_{\ell}^{m}(\theta, \phi),
\label{r2}.
\end{equation}
where $Y_{\ell}^{m}(\theta, \phi)$ is the spherical harmonic function, the radial Klein-Gordon equation for $\ell = 0$ is obtained as 
\begin{equation}
\hbar^{2} c^{2} \frac{d^{2} U_{n, \ell}(r)}{dr^{2}}  + \left \{ \left[ E_{R}^{2} - \mu^{2} c^{4}\right] -  \left[ \mu c^{2}V_{s}(r) + E_{R} V_{s}(r)\right] + \left[ \frac{V_{v}^{2}(r)}{4} -\frac{V_{s}^{2}(r)}{4} \right] \right\}U_{n, \ell}(r) = 0
\label{r3}.
\end{equation}

We consider the case when the scalar and vector potentials are equal to a q-deformed hyperbolic modified P\"{o}schl-Teller potential model, which is expressed as (P\"{o}schl and Teller 1933, Landau and Liftshitz 1977, Fl\"{u}gge 1994, Grosche and Steiner 1998, E$\check{g}$rifes et al. 1999, Dong and Dong 2002, Grosche 2005, Zhao et al. 2005) 
\begin{equation}
V(r) = - \frac{D}{\cosh_{q}^{2}(\alpha r)}
\label{r4}.
\end{equation}
In Figure $1$, we show the radial as well as the deformation dependence of the deformed modified P\"{o}schl-Teller potential.  On the graph, blue line is the graph of $q = 1$; green line is the graph of $q = 2$; red line is the graph of $q = 3$; dark line is graph of $q = 4$; short dash dark line is the graph of $q = 5$ and long dash dark line is the graph of $q = 6$.

Using the definitions of the deformed hyperbolic functions (Arai 1991), we have:
\begin{eqnarray}
&\sinh_{q}x = \frac{1}{2}(e^{x} - qe^{-x}),~\cosh_{q}x = \frac{1}{2}(e^{x} + qe^{-x}),~ \tanh_{q}x = \frac{\sinh_{q}x}{\cosh_{q}x},\\ \nonumber 
& \coth_{q}x = \frac{1}{\tanh_{q}x}
\label{r5},
\end{eqnarray}
We have, 
\begin{eqnarray}
&\cosh_{q}^{2}x - \sinh_{q}^{2}x = q,~ \frac{d}{dx}\cosh_{q}x = \sinh_{q}x,~ \frac{d}{dx}\sinh_{q}x = \cosh_{q}x  \\ \nonumber
&\frac{d}{dx}\tanh_{q}x = \frac{q}{\cosh_{q}^{2}x}, ~\frac{d}{dx}\coth_{q}x = -\frac{q}{\sinh_{q}^{2}x}
\label{r6},
\end{eqnarray}
 where $q>0$  is a real parameter.  When $q$ is complex, the above deformed hyperbolic functions are called the generalized deformed (q-deformed) hyperbolic functions (E$\check{g}$rifes et al. 1999, Yi et al. 2004, Grosche 2005, Zhao et al. 2005).  

	With equation (\ref{r4}), equation (\ref{r3}) becomes
\begin{equation}
U_{n,~\ell}^{~''}(r) + \left[\frac{(E_{R}^{2} - \mu^{2}c^{4})}{\hbar^{2}c^{2}} +  \frac{D(\mu c^{2} + E_{R})}{\hbar^{2}c^{2}\cosh_{q}^{2}(\alpha r)}\right]U_{n,~\ell}= 0
\label{r7}.
\end{equation}	
Introducing the parameters  $\tilde{E_{R}^{2}} = \frac{(E_{R}^{2} - \mu^{2}c^{4})}{\hbar^{2}c^{2}}$, $ \gamma(\gamma + 1) = \frac{D(\mu c^{2} + E_{R})}{ \alpha^{2} \hbar^{2}c^{2}} $    and by using a new variable $y = \tanh_{q}(\alpha r)$, equation (\ref{r7}) becomes
\begin{equation}
\frac{d}{dy}\left[ (1 - y^{2})\frac{d}{dy}U_{n, ~\ell}(y)\right] + \left[\frac{\gamma(\gamma + 1)}{q} - \frac{\xi^{2}}{1 - y^{2}} \right]U_{n, ~\ell}(y) = 0
\label{r8},
\end{equation}
where 
\begin{equation}
\xi = \sqrt{\frac{- \tilde{E_{R}^{2}} }{\alpha^{2}}}
\label{r9}.
\end{equation} 	 										
We seek for the exact solution of equation (\ref{r8}) via the following ansatz:
\begin{equation}
U_{n, ~\ell}(y) = (1 - y^{2})^{\frac{\xi}{2}}f(y)
\label{r10},
\end{equation} 	 										
with equation (\ref{r10}), equation (\ref{r8}) becomes
\begin{equation}
(1 - y^{2}) f''(y) - 2y(\xi + 1)f'(y) + \left[\frac{\gamma(\gamma + 1)}{q} + \frac{\xi(\xi - 2)y^{2}}{(1 - y^{2})} - \xi + \frac{2 \xi y^{2}}{(1 - y^{2})} \right]f(y) = 0
\label{r11}.
\end{equation}
By defining the variable y as
\begin{equation}
\eta = \frac{1}{2}(1- y)
\label{r12},
\end{equation}
then, equation (\ref{r11}) can be rewritten as
\begin{equation}
\eta(1 - \eta) f''(\eta) + \left[(\xi + 1) - 2 \eta(\xi + 1) \right]f'(\eta) + \left[\frac{\gamma(\gamma + 1)}{q} -  \xi(\xi + 1) \right]f(\eta) = 0
\label{r13}.
\end{equation}
Equation (\ref{r13}) is the hypergeometric differential equation and has the solutions (Abramowitz and Stegun 1970, Arfken and Weber 1994, Gradsyteyn and Ryzhik 1994, Andrew et al. 2000):
\begin{equation}
\Psi_{n}^{q}(y) = N^{q}(1 - y^{2})^{\frac{\xi}{2}}~ _{2}F_{1} \left(-n,~-n + 2k, ~-n + k + \frac{1}{2};~ \frac{1 - y}{2}\right)
\label{r14},
\end{equation}
where $ _{2}F_{1} \left(-n,~-n + 2k, ~-n + k + \frac{1}{2};~ \frac{1 - y}{2}\right)$ are the hypergeometric polynomials of degree $n$ and $k = \sqrt{\frac{1}{4} + \frac{\gamma (\gamma + 1)}{q}}$.

	We next consider the relationship between the hypergeometric functions and the Gegenbauer polynomials (Abramowitz and Stegun 1970, Arfken and Weber 1994, Gradsyteyn and Ryzhik 1994, Andrew et al. 2000), that is,                     
\begin{equation}
C_{n}^{\lambda}(x) = \frac{\Gamma{(2\lambda + n)}}{n! \Gamma{2 \lambda}}  ~ _{2}F_{1} \left(-n,~-n + 2\lambda, ~\lambda + \frac{1}{2};~ \frac{1 - x}{2}\right)
\label{r15},
\end{equation}
On writing equation (\ref{r14}) in terms of the Gegenbauer functions, we have
\begin{equation}
\Psi_{n}^{q}(y) = N^{q}(1 - y^{2})^{\frac{k - n - \frac{1}{2}}{2}} C_{n}^{~ -n + k}(x) 
\label{r16},
\end{equation}
where $N^{q}$ is the normalization constant to be determined from the normalization condition
\begin{equation}
\int_{-\infty}^{\infty} [\Psi_{n}^{q}(x)]^{2}dx = \frac{[N^{q}]^{2}}{\alpha} \int_{-1}^{1} (1 - y^{2})^{k - n - \frac{1}{2}} [ C_{n}^{~ -n + k}(y)]^{2}dy =1 
\label{r17}.
\end{equation}
The integrand in equation (\ref{r17}) can be evaluated by using the integrals (Abramowitz and Stegun 1970, Arfken and Weber 1994, Gradsyteyn and Ryzhik 1994, Andrew et al. 2000, Dong and Dong 2002, Dong 2007):
\begin{equation}
\int_{-1}^{1}  (1 - x)^{\nu - \frac{3}{2}} (1 + x)^{\nu - \frac{1}{2}} [ C_{n}^{ \nu}(x)]^{2}dx  = \frac{\pi^{1/2} \Gamma{(\nu - \frac{1}{2})} \Gamma{(2 \nu + n)}}  { n! \Gamma{(\nu )} \Gamma{(2 \nu)}}; ~~Re~ \nu~>1 
\label{r18},
\end{equation}            
since
\begin{eqnarray}
\int_{-1}^{1}  (1 - x)^{\nu - \frac{3}{2}} (1 + x)^{\nu - \frac{1}{2}} [ C_{n}^{ \nu}(x)]^{2}dx 
 = \int_{-1}^{1}  (1 - x^{2})^{\nu - \frac{3}{2}} (1 + x) [ C_{n}^{ \nu}(x)]^{2}dx \\ \nonumber
= \int_{-1}^{1}  (1 - x^{2})^{\nu - \frac{3}{2}} [ C_{n}^{ \nu}(x)]^{2}dx  +  \int_{-1}^{1}  (1 - x^{2})^{\nu - \frac{3}{2}} x [ C_{n}^{ \nu}(x)]^{2}dx \\ \nonumber 
 =\frac{\pi^{1/2} \Gamma{(\nu - \frac{1}{2})} \Gamma{(2 \nu + n)}}  { n!\Gamma{(\nu )} \Gamma{(2 \nu)}}
\label{r19}.
\end{eqnarray}            
Note that $\int_{-1}^{1}  (1 - x^{2})^{\nu - \frac{3}{2}} x [ C_{n}^{ \nu}(x)]^{2}dx = 0$, due to the odd parity of the integrand. Then, $N^{q}$, which is the normalization constant in equation (\ref{r17}), gives
\begin{equation}
N^{q} =\sqrt{\frac{\alpha n!(k - n - 1)!(2k - 2n - 1)!  }{\pi^{1/2} (k - n - \frac{3}{2})!(2k - n - 1)!  }   }  
\label{r20}.
\end{equation}                  		  							
Therefore, the wave function of the Klein-Gordon equation for the q-deformed hyperbolic modified P\"{o}schl-
Teller potential in terms of the Gegenbauer functions is
\begin{equation}
\Psi_{n}^{q}(r) = N^{q} ~~ q^{\frac{k - n - \frac{1}{2}}{2}} [\sec h_{q}^{2}(\alpha r)]^{\frac{k - n - \frac{1}{2}}{2}} C_{n}^{~ -n + k}(\tanh_{q}(\alpha r)) 
\label{r21},
\end{equation}
where $N^{q}$ is as given in equation (\ref{r20}). The corresponding relativistic bound state energy spectra are obtained from equation (\ref{r9}) as
\begin{equation}
E_{R}^{2} - \mu^{2} c^{4} = - \alpha^{2} \hbar^{2} c^{2}\left[ \left(n + \frac{1}{2}\right) -k \right]^{2}; n = 0, 1, 2, \ldots 
\label{r22},
\end{equation}
where $k = \sqrt{\frac{1}{4} + \frac{\gamma (\gamma + 1)}{q} }$.

\section{Discussions}

In this section, in the framework of the Klein-Gordon theory with equal scalar and vector potentials, the relativistic bound state energy spectra and the corresponding wave functions for the reflectionless-type potential, q-deformed symmetric hyperbolic modified P\"{o}schl-Teller potential, symmetric modified P\"{o}schl-Teller potential and the PT-symmetric version of the  hyperbolic modified P\"{o}schl-Teller potential are obtained by choosing appropriate parameters in the q-deformed hyperbolic modified P\"{o}schl-Teller potential.

\subsection{Reflectionless-type potential}
On putting $q = 1$, $\alpha = 1$ and $D = \frac{1}{2} \lambda(\lambda + 1)$ in the q-deformed hyperbolic modified P\"{o}schl-Teller potential, then, equation (\ref{r4}) reduces to the reflectionless-type potential (Zhao et al. 2005, Setare and Haidari 2010). 
\begin{equation}
V(r) = -\frac{1}{2}\lambda (\lambda + 1)\sec h^{2}r
\label{r23},
\end{equation}
where $\lambda$ is an integer, the wave function and the relativistic bound state energy spectra for the Klein-
Gordon equation with equal scalar and vector reflectionless-type potential are respectively obtained as:
\begin{equation}
\Psi_{n}^{RT}(r) = N^{RT} ~~  [\sec h^{2}(r)]^{\frac{k - n - \frac{1}{2}}{2}} C_{n}^{~ -n + k}(\tanh( r)) 
\label{r24},
\end{equation}
and 
\begin{equation}
E_{RT}^{2} - \mu^{2} c^{4} = - \hbar^{2} c^{2}\left[ \left(n + \frac{1}{2}\right) -k \right]^{2}; n = 0, 1, 2, \ldots 
\label{r25},
\end{equation}
where $k = \sqrt{\frac{1}{4} + \frac{\lambda (\lambda + 1)(E_{RT} + \mu c^{2})}{2\hbar^{2} c^{2}} }$ and  $N^{RT} =\sqrt{\frac{ n!(k - n - 1)!(2k - 2n - 1)!  }{\pi^{1/2} (k - n - \frac{3}{2})!(2k - n - 1)!  }   }  $~~.  

\subsection{q-deformed symmetric hyperbolic Modified P\"{o}schl-Teller potential}
Choosing $\alpha = 1$ and $D = \lambda^{2} - \frac{1}{4}$ in equation (\ref{r4}) (Grosche 2005), we have 
\begin{equation}
V(r) = -\frac{\lambda^{2} - \frac{1}{4}}{\cosh_{q}^{2}(r)}
\label{r26},
\end{equation}
again, the relativistic bound state energy spectra and the wave function for the Klein-Gordon equation with equal scalar and vector symmetric potential are :
\begin{equation}
E_{qs}^{2} - \mu^{2} c^{4} = - \hbar^{2} c^{2}\left[ \left(n + \frac{1}{2}\right) -k \right]^{2}; n = 0, 1, 2, \ldots 
\label{r27},
\end{equation}
and 
\begin{equation}
\Psi_{n}^{qs}(r) = N^{qs} ~~ q^{\frac{k - n - \frac{1}{2}}{2}} [\sec h_{q}^{2}(r)]^{\frac{k - n - \frac{1}{2}}{2}} C_{n}^{~ -n + k}(\tanh_{q}(r)) 
\label{r28},
\end{equation}
where $k = \sqrt{\frac{1}{4} + \frac{(\lambda^{2} - \frac{1}{4})(E_{qs} + \mu c^{2})}{ q \hbar^{2} c^{2}} }$ and  $N^{qs} =\sqrt{\frac{ n!(k - n - 1)!(2k - 2n - 1)!  }{\pi^{1/2} (k - n - \frac{3}{2})!(2k - n - 1)!  }   }  $~~.  

\subsection{Symmetric hyperbolic modified P\"{o}schl-Teller potential}
Choosing $q = 1$, $\alpha = 1$ and $D = \lambda^{2} - \frac{1}{4}$ in the q-deformed hyperbolic modified P\"{o}schl-Teller potential (Grosche and Steiner 1998, Oyewumi and Bangudu 1999, 2000, Grypeos et al. 2004, Oyewumi et al. 2004, Efthymiou et al. 2008, Oyewumi 2009), equation (\ref{r4}) becomes 
\begin{equation}
V(r) = -\frac{\lambda^{2} - \frac{1}{4}}{\cosh^{2}(r)}
\label{r29},
\end{equation}
with the following solutions:
\begin{equation}
E_{s}^{2} - \mu^{2} c^{4} = - \hbar^{2} c^{2}\left[ \left(n + \frac{1}{2}\right) -k \right]^{2}; n = 0, 1, 2, \ldots 
\label{r30},
\end{equation}
and 
\begin{equation}
\Psi_{n}^{s}(r) = N^{s} [\sec h^{2}(r)]^{\frac{k - n - \frac{1}{2}}{2}} C_{n}^{~ -n + k}(\tanh(r)) 
\label{r31},
\end{equation}
where $k = \sqrt{\frac{1}{4} + \frac{(\lambda^{2} - \frac{1}{4})(E_{s} + \mu c^{2})}{  \hbar^{2} c^{2}} }$ and  $N^{s} =\sqrt{\frac{ n!(k - n - 1)!(2k - 2n - 1)!  }{\pi^{1/2} (k - n - \frac{3}{2})!(2k - n - 1)!  }   }  $~~.  

\subsection{PT-Symmetric version of the hyperbolic modified  P\"{o}schl-Teller potential}
If we substitute $D = - Dq_{c}$, $q = - q_{c}$ and $q_{c} = e^{2 i \alpha \epsilon}$, then, potential in equation (\ref{r4}) becomes the PT-symmetric version of the hyperbolic modified P\"{o}schl-Teller potential, here $D>0$, $|\epsilon| > \pi/4$, that is,
\begin{equation}
V(r) = \frac{Dq_{c}}{\frac{1}{4} [e^{2 \alpha r} - 2q_{c} + q_{c}^{2}e^{-2 \alpha r}] } =  Dq_{c} cosec{h_{q_{c}}^{2}}
(\alpha r)
\label{r32}.
\end{equation}
With these substitutions, it follows that the relativistic bound state energy spectra and the corresponding wave functions are respectively obtained as:
\begin{equation}
E_{q_{c}}^{2} - \mu^{2} c^{4} = - \alpha \hbar^{2} c^{2}\left[ \left(n + \frac{1}{2}\right) -k \right]^{2}; n = 0, 1, 2, \ldots 
\label{r33},
\end{equation}
and 
\begin{equation}
\Psi_{n}^{q_{c}}(r) = N^{q_{c}}~~ (-q_{c})^{\frac{k - n - \frac{1}{2}}{2}}[cosec h_{q_{c}}^{2}(\alpha r)]^{\frac{k - n - \frac{1}{2}}{2}} C_{n}^{~ -n + k}(\coth_{q_{c}}(\alpha r)) 
\label{r34},
\end{equation}
where $k = \sqrt{\frac{1}{4} +  \frac{D(E_{q_{c}} + \mu c^{2})}{ \alpha \hbar^{2} c^{2}} }$ and  $N^{q_{c}} =\sqrt{\frac{ \alpha n!(k - n - 1)!(2k - 2n - 1)!  }{\pi^{1/2} (k - n - \frac{3}{2})!(2k - n - 1)!  }   }  $~~.  

\section{Conclusion}
In this research work, we have investigated the s-wave bound states of the Klein-Gordon equation with equal scalar and vector q-deformed hyperbolic P\"{o}schl-Teller potential. The energy eigenvalue equation and the normalized radial wave function are obtained analytically. The radial wave functions of spin $0$ particles are expressed in terms of the Gegenbauer functions by exploiting the relationship between the Gegenbauer functions and the hypergeometric functions.  
The energy equations and the corresponding wave functions of the reflectionless-type potential, the q-deformed symmetric hyperbolic P\"{o}schl-Teller potential, the symmetric P\"{o}schl-Teller potential, and the PT-symmetric version of the hyperbolic P\"{o}schl-Teller potential are obtained as special cases of  the q-deformed hyperbolic P\"{o}schl-Teller potential in the Klein-Gordon theory with equally mixed scalar and vector potentials.

\vspace{1cm}

{\bf Aknowledgements}

\noindent
The authors are indebted to the following people for the motivations received from them: Emeritus Prof.  K. T. Hecht (USA), Prof. S. H. Dong (Mexico), Prof. C. Berkdermir (Turkey), Prof. C. S. Jia (China), Prof. J. L. 
L$\grave{o}$pez-Bonila  (Mexico), Prof. Dr. C. Grosche (Germany), Prof. M. N. Hounkonnou (Benin Republic), Prof. J. Govaerts (Belgium) and Prof. W. C.  Qiang (China).  Also, the authors thank the anonymous kind referees and editors for the constructive comments and suggestions.

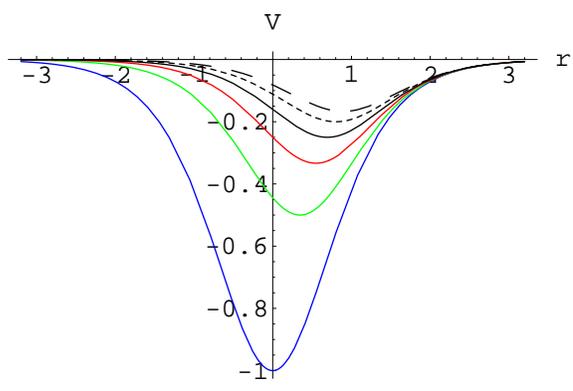
\end{document}